# A confocal laser-induced fluorescence diagnostic with a ring-shaped laser beam

I. Romadanov, Y. Raitses

*Princeton Plasma Physics Laboratory, Princeton, NJ 08543, USA*

**ABSTRACT**

In this work, we report a confocal laser induced fluorescence (LIF) configuration, which allows for high spatial resolution measurements of plasma properties in plasma setups and sources with a limited optical access. The proposed LIF configuration utilizes a ring-shaped laser beam generated by a pair of diffractive axicons. LIF signal is collected along the main optical axis within the ring region. It is shown experimentally that at the focal distance of 300 mm, the spatial resolution of approximately 5.3 mm can be achieved. More than that, theoretical estimations predict the possibility of achieving ~1 mm resolution at the same focal distance by adjusting laser beam parameters. This is approaching the localization accuracy of conventional LIF collection methods (with crossing laser beam injection and fluorescence collection optical paths). Measurements of the ion velocity distribution function in an argon plasma using both the confocal LIF with ring shaped laser beam and conventional LIF demonstrate a satisfactory agreement. The proposed LIF setup has potential applications for diagnostics in various plasma processing equipment, and plasma sources such as hollow cathodes, microplasmas, electric propulsion, etc.

## I. INTRODUCTION

Laser induced fluorescence (LIF) is a non-invasive, active optical diagnostic technique that is commonly used for measurements of velocity distribution functions (VDF) of ion or neutral species in low temperature plasmas in a range of conditions, from weakly collisional low-pressure plasmas to collisional plasmas generated at elevated pressures. The principle of LIF diagnostic is based on optical pumping of a plasma sample with a laser light and analyzing the resulting fluorescence signal. By using a narrow linewidth tunable laser, this diagnostic method is able to provide high-resolution information about the velocity distribution function of plasma species [1]. This makes LIF a versatile and valuable tool for plasma physicists and engineers studying plasma behavior in various applications. The VDF obtained from LIF can provide information about important parameters such as ions/atoms temperatures and flow velocities. When the relation between the ground state and excited states are known, the absolute densities can be deduced from VDF data. In this paper, we focus on LIF measurements with excitation from metastable levels in weakly collisional plasma with non-equilibrium VDFs. Under such conditions, the measurements of the LIF signal, as laser wavelength is scanned over Doppler broadened transition, provide a profile of the VDF. This approach is commonly used for measurements of VDFs for ions and atoms of numerous gases with applications to, for example, plasma thrusters, ion sources, and basic plasma science experiments such as plasma-wall interactions, double layers, etc. [2–11]

Conventional LIF diagnostic requires optical access to plasma from two direction: one is for the laser beam injection, and another is for the fluorescence emission collection [12]. In such configuration, spatial resolution is defined as an overlap between injection and collection focal regions. This overlap is minimized when an angle between the laser beam and the collection optical axis is $90^o$, resulting in sub mm spatial resolutions. However, this sets a limit on the applicability of these technique for some plasma sources, as it is not always possible or allowable to provide the required optical access (e.g., plasma hollow cathodes and anodes, plasma processing reactors for microelectronics). Several in-depths reviews of conventional LIF measurements with intersecting optical paths are available in the literature. [13,14]

Confocal configuration of laser induced diagnostics is widely used in biology and medicine,[15–17] and there are several works demonstrating its application for Raman spectroscopy,[18,19] Thompson scattering,[20] and for laser induced fluorescence diagnostic. [21,22] Its main advantage is that the laser beam injection and fluorescence collection branches coincide. However, the spatial resolution of confocal system is mainly defined by the depth of field (DOF) of the focusing/collection optics, and it is difficult to maintain DOF small with the increasing distance to the measurements point.

Confocal optical arrangement is used for two-photon LIF (TALIF), where spatial resolution is defined by DOF, and the fact that TALIF intensity is proportional to the squared laser beam intensity. Therefore, measurements are highly localized within the DOF. In the works [22,23] a coaxial TALIF system was designed, where the collection lens has an aperture at the center, which is used for passing the laser beam. Fluorescence signal is collected from the hollow cone, which base is defined by the difference between lens and aperture diameters. The reported spatial resolution was about 20 mm at $f = 820\ mm$ focal distance. In more resent work [24] the conservative estimation for TALIF measurements resolution of 5 mm was reported.

For single photon LIF, the fluorescence intensity is directly proportional to the laser beam intensity, thus, spatial resolution is solely defined by the DOF of the optical setup. In the number of previous works [25,26], there were attempts to overcome this issue, especially for setups with the long focal distances. In work [21], 3 cm at 20 cm focal length spatial resolution was achieved by using the fact that DOF is inversely proportional to the focused laser beam diameter. Thus, the laser beam was expanded in a beam expander and then, focused with 5 cm lens. In [25], confocal design included the obstruction disk, which cuts the fluorescence light along the optical axis. Spatial resolution is then defined by overlap region between the hollow fluorescence cone and the laser beam. The reported resolution was about 1.4 mm for $f = 150\ mm$ and ~7.3 mm at $f = 500\ mm$ focal distance. [26]

In all the above LIF configurations, the laser beam remains the same cylindrical shape and the spatial resolution was ensured either by expanding the laser beam or by cutting the fluorescence light along the laser beam, thus reducing the overlap between the injection and collection paths. While in case of the expanded beam the collected fluorescence light was not cut the achievable spatial resolution is moderate. Cutting the fluorescence light along the optical axis would result in reduction of the signal-to-noise ratio (SNR), which can be crucial for the cases of low plasma density or bright emitting background within the field of view (e.g., thermionic cathode).

Here, we present a proof-of-concept of LIF configuration with a ring-shaped laser beam, which allows to ensure a relatively high spatial localization of a few mm's over large focal distances of >200 mm with a high SNR (>10). The ring-shaped laser beam of large diameter (22 mm) ensures small DOF. All fluorescence light enclosed by the focused laser beam is collected, thus, resulting in high SNR. A set of diffractive axicon lenses were used to create ring-shaped laser beam of high quality. With the current setup the achieved spatial resolution is 5.3 mm at $f = 500\ mm$ focal distance. Here, the spatial resolution is defined as a region with 95% of focused laser beam intensity. The paper is organized as follows. Experiment setup and LIF diagnostics used in this work are described in Section 1 and 2, respectively. Measurement results are presented and discussed in Section 3 followed by conclusions.

## II. EXPERIMENTAL SETUP

The reported experiments were performed with an argon plasma generated by a 100 W DC discharge with a thermionic cathode in a fully enclosed container with an approximate volume of $10^5$ mm$^3$ (see FIG. 1). The source is installed in the vacuum chamber made from a standard 10" diameter six-way cross. The chamber is equipped with mechanical and turbomolecular pumps. The source has a 1 cm diameter opening on the longer front wall (FIG. 1). This opening allows vacuum pumping of the gas from the source and access for diagnostics of the enclosed plasma (e.g., by probes, optical emission spectroscopy and laser diagnostics). In the described experiments, the argon background gas pressure measured with a commercial ion gauge in the vacuum chamber was in a 100 microtorr range, while the argon gas pressure in the source was estimated to be about ten times higher. Under such conditions, plasma in the source is weakly collisional with the ion mean free path, $\frac{\lambda_i}{L} \gg 1$, while plasma leaking through the opening is nearly collisionless. Here, L is the width of the source in z-direction.

For the implementation of the confocal and conventional LIF configurations, the laser beam with the wavevector $\overline{k}$ and frequency $\nu$ is injected through the 10 mm opening on the front wall. In addition, for the conventional LIF, the source has additional three 3 mm diameter orifices between on its side wall. These holes are used to collect LIF signal from the three spatial locations (Z = 0.0 to Z = 1.0) as show in FIG. 1.

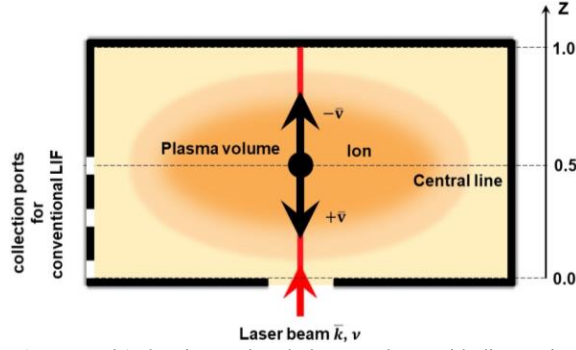

FIG. 1. Schematic of the plasma source (not to scale) showing enclosed plasma volume with diagnostic opening of about 10 mm diameter at the front side and three collection ports of 3 mm diameter each on the side, which were used for conventional LIF. Confocal LIF collection points were evenly spread between Z = 0.0 to Z = 1.0 (dimensionless locations).

In this work, LIF measurements are performed by sweeping the frequency of a tunable diode laser with a narrow linewidth over the absorption line of an argon ion which is broadened due to Doppler shift. By measuring of the emission intensity at different laser wavelengths a VDF profile can be recovered. For an argon ion, a three-state scheme is used. Selected transition is illustrated in FIG. 2.

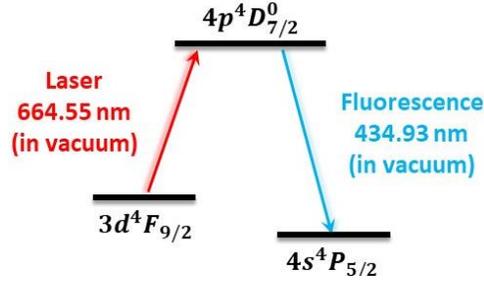

FIG. 2. LIF transition for Ar-II (argon ion).

Ar ion at $3d^4F_{9/2}$ metastable state is optically pumped by 664.553 nm (in vacuum) laser light to $4p^4D_{7/2}$ state, which decays to $4s^4P_{5/2}$ state by an emission at 434.929 nm [27,28]. This transition was chosen over more commonly used $3d^4F_{7/2} - 4p^4D_{5/2}$ (excitation at 668.614 nm, emission at 442.724 nm) [29] as having better signal intensity for given experimental conditions. Plasma ion density profiles along $z-$axis was measured with the biased Langmuir probe [30] installed on a moveable stage. Distance increment was set to 2 mm. These profiles were used to compare with metastable density profiles obtained from VDF data.

## III. CONFOCAL LIF SETUP WITH ANNULAR LASER BEAM

### A. Annular beam generation with axicons

Axicon is an optical element, which allows for annular beam formation [31]. The principle of operation of the refractive axicon is shown in FIG. 3a. Due to conical shape of the axicon the passing laser beam with the diameter $2\delta$ is transformed into the ring-shaped beam with the annulus thickness of $\approx \delta$. Such beam can be collimated with the second axicon of the same geometry. Diameter of collimated beam is determined by the distance between two axicons.

Refractive axicons [32], as shown in FIG. 3b, suffer from beam interference due to imperfect geometry of the optical element. This results in formation of the region with the residual intensity within the laser light ring (see Ref. 6 for more details). This will result in the reduction of the spatial resolution, due to increased focal spot size.

In this work a pair of diffractive axicons (DA), which were custom made by HOLO/OR for 670 nm wavelength, were used [33]. Diffractive axicon results in very low residual intensities of the central region, as shown in FIG. 3c. The quality of the annular beam is sensitive to the input beam shape and beam collimation. To improve the performance of the DA element the laser beam was converted into the circular beam with Gaussian intensity profile by fiber coupling and follow-up collimation (see Section 2.3).

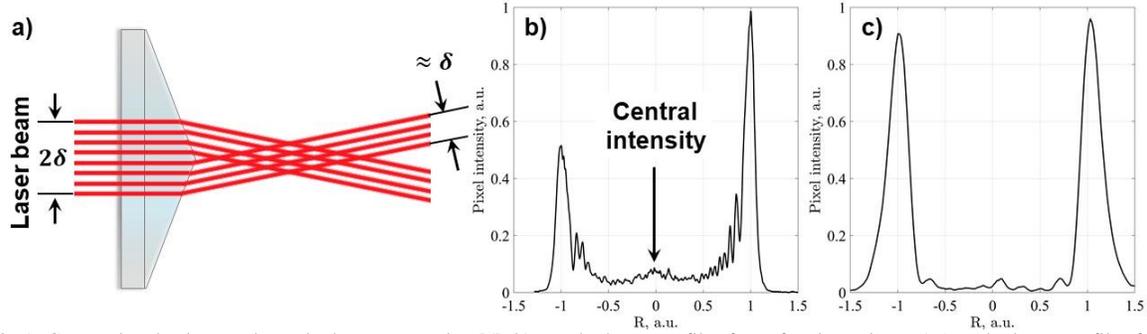

FIG. 3. a): Conventional axicon and annular beam generation [5]; b) annular beam profile after refractive axicon; c) Annular beam profile after the DA.

### B. Light collection in the confocal setup with the annular beam

Light collection with the proposed confocal setup is illustrated in FIG. 3. Collimated annular laser beam is focused with the lens $L$. Fluorescence intensity is maximized in the region with maximum laser intensity at the focal distance $f$. Fluorescence light, which is spread in all direction from the focal point, is collected with the same lens $L$, forming the light cone with a base $L$ and height $f$ (shown in blue in FIG. 4, not to scale). The base of this cone can be reduced, so it is not overlapped with the laser beam path (see section 3.3). From this sketch one can see that the spatial resolution is defined only by the overlap volume between the laser beam, and collection cones. The contour of the overlap area is marked with black lines. However, due to nonuniform laser intensity distribution, only fluorescence light from the shaded area is a main contributor to the detected signal.

As it will be shown in section 3.4 that overlap volume can be controlled by the ring thickness $\delta$ and annulus radius $R$. All fluorescence light, that is within the laser beam cone, can be collected without loss of the spatial resolution. This is the main difference of the suggested setup as compared to Refs [25,26], where spatial resolution is controlled by the obstruction disk, and higher spatial resolution can be achieved only by reduction of the collected fluorescence light, which leads to SNR reduction.

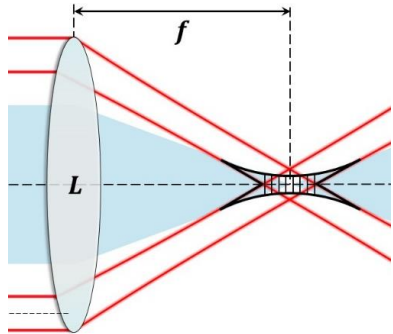

FIG. 4. Fluorescence light collection with the proposed confocal design. L is a lens with the focal distance f. Red lines mark the contour of the annular laser beam, blue region is the collected cone of the fluorescence light. Black line is the intersection of the fluorescence light cone and the laser beam hollow cone. Black shaded area is the collection volume.

### C. Laser induced fluorescence setup

The LIF system is built around single-mode TOptica DLC DL PRO 670 tunable diode laser (TDL) and its design is shown in FIG. 5. The diode laser is Littrow-type grating stabilized external cavity design and it has coarse tuning range between 660 to 673 nm and mode-hop free tuning range of 20 GHz. The output power is wavelength dependent and maximum value is about 23 mW. Short term wavelength stability is 600 kHz (over 5 $\mu s$). The laser wavelength is controlled by scanning the voltage applied to the piezo actuator.

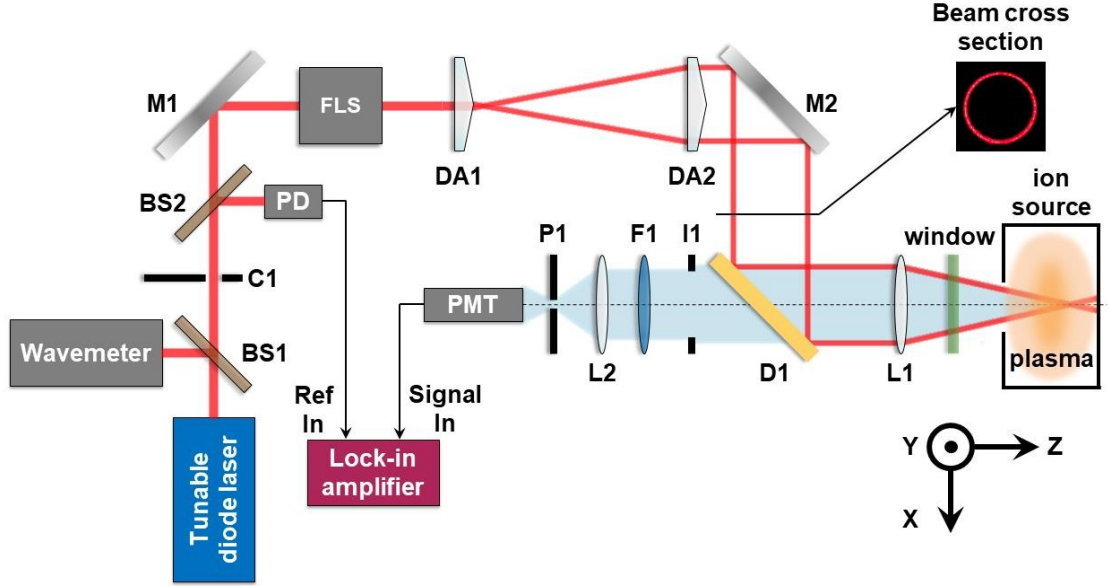

FIG. 5. Block diagram of the LIF setup and beam path into the ion source. BS – beam splitter; C – mechanical chopper, PD – photodiode; M – mirror; FLS – fiber launch system; DA – diffractive axicon; L – lens; D – dichroic mirror; I – iris; F – bandpass filter; P – pinhole; PMT - Photomultiplier tube.

Part of the laser beam is sampled by beam splitter (BS1) and is directed to the Bristol 621-A wavemeter, which has a built-in continuous calibration (single-frequency He–Ne laser, accuracy ±0.0002 nm at 1000 nm). With this accuracy the achievable velocity resolution is ~60 m/s. Main beam is modulated by the mechanical chopper (C1), which is set to a frequency of 2 kHz. Modulated beam is sampled with the beam splitter (BS2) and the sampled beam is set to the photodiode (PD). Output of the photodiode provides the reference frequency for the lock-in amplifier.

Beam is then directed by the mirror M1 through the fiber launch system (FLS) Thorlabs MBT613D into a single mode fiber. This is to ensure the circular beam shape with Gaussian profile. The fiber output was collimated with F260FC-B collimator providing 3 mm diameter beam.

Beam is then passed through a pair of diffractive axicon lenses (DA1, 2 - HOLO/OR 670 nm), which are installed in XY adjustable mounts to ensure their coaxiality. Both diffractive axicons are installed on rotation (around Y axis) stages to ensure that they are parallel to each other. Beam quality after axicons was verified with the CMOS camera. Circularity, equal intensity distribution, and absence of light intensity within the ring were monitored. After axicons, annular beam (with diameter of ~22 mm) is directed by mirror M2 and dichroic mirror D1 (shortpass, 490 nm cutoff) into the vacuum chamber. Beam focusing is achieved by lens L1 with the focal distance $f = 300$ mm.

The fluorescence light is collected by the same lens L1 and the passed through the dichroic mirror D1. The outer part of the collimated light cylinder is then cut by iris I1. This is to cut part of the fluorescent light from species, which are excited along the laser beam path. Collected light is then passed through the bandpass filter 430/10 nm (Thorlabs FBH430-10) and then focused by lens L2 into the $100 \mu m$ pinhole P1. After the pinhole the light is sent into photomultiplier tube PMT. PMT output is fed into the lock-in amplifier input and fundamental component arising due to laser beam amplitude modulation is filtered out. Laser wavelength, and readings from wavemeter and lock-in amplifier outputs (signal and phase) are collected with in-house developed LabVIEW based software.

The fluorescence excitation line shape can broaden if transition is saturated due to high laser intensity [34]. LIF signal intensity was measured as a function of the laser power to ensure such line-shaped distortions are prevented. Linear relation between fluorescence signal and laser power was obtained.

### D. Spatial resolution and its characterization

In this section a simplified model for spatial resolution estimation is presented. Spatial resolution or depth of field (DOF) is defined by the intersection between annular beam paths at the focal point. This is illustrated in Fig. 6, where annular beam of radius $R$ and ring thickness $\delta$ is focused by lens with a focus length $f$. From geometrical considerations the length of DOF can be found as

$$DOF = \frac{4\lambda f}{\pi \delta} \frac{1}{\sin(\operatorname{atan}(R/f))}, \qquad (1)$$

where $\lambda$ is the laser wavelength (see Eq. 4 in Ref. [35]). For the setup used is this work, $R \approx 11\ mm$, $\delta = 1.5\ mm$, $f = 300mm$, $\lambda = 668nm$ the theoretical DOF is approximately 4.6 mm. Note that this is an upper limit due to the laser intensity being not uniformly distributed (Gaussian) along the DOF. Such nonuniformity will result in smaller effective spatial resolution. However, it is important to notice that this is simplified model, which doesn't consider hyperbolic shape of focused laser beam and the effect of full volume of the overlap between the injection and the collection paths.

From the Eq. (1) one can see that at fixed focal length $f$ the DOF can be minimized by increasing $\delta$ or annulus radius $R$. For example, for annular beam with $R = 20mm$ and $\delta = 3mm$, it should be possible to achieve ~1 mm resolution. In the current setup the main limit for the spatial resolution improvement is the axion diameters, which is 25.4 mm and clear aperture is about 22 mm. By utilizing 2-inch optics (with clear aperture of ~48 mm) and further increasing ring thickness $\delta$ to 4mm (with commercially available fiber collimators) it is possible to achieve spatial resolution of 0.8 mm at $f = 300mm$.

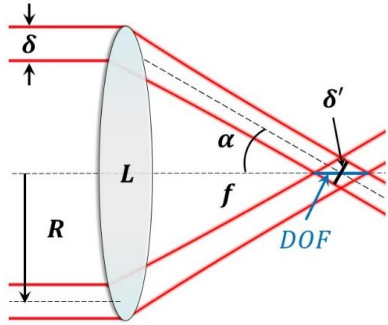

FIG. 6. Diagram showing DOF for the annular beam. R is annular beam radius; δ is ring thickness; L is lens with a focal length f; δ' is the beam waist at the focal plane; DOF is the depth of focus.

To experimentally characterize the spatial resolution and the optical response function of our system, the CMOS camera on the movable stage was installed at the focal point of the lens L1 (see FIG. 4). The schematic of the experiment is shown in FIG. 7a. Camera was moved along the optical axis (Z) and beam profile images were collected. Origin was selected at the lens focal point. Camera exposure was adjusted at each position to avoid pixels saturation. All images were stacked together allowing for the laser beam profile visualization (see FIG. 6b).

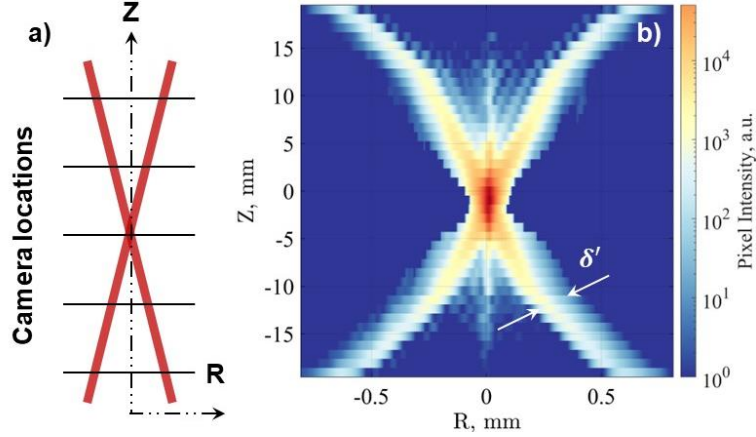

FIG. 7. Spatial resolution characterization. a) Schematic of experiment for spatial resolution characterization; b) Cross section of the beam profile, where δ′ is the beam thickness at the lens focal plane.

Spatial resolution and optical response function were determined from the intensity profiles along the centerline at $R = 0mm$, which is shown in Fig. 8. Conservatively, resolution was defined as a region which contains 95% of the laser beam intensity (blue line in Fig. 8). Experimentally found value of 5.3 mm is close to the theoretical value of 4.6 mm obtained from Eq. (1). Resolution of the conventional method, defined as a region with 99.73% intensity is about 1.2 mm. If spatial resolution is defined as a region where only 68% of the laser beam intensity is contained, then the resolution of the confocal method is 2.5 mm (red line in Fig. 8).

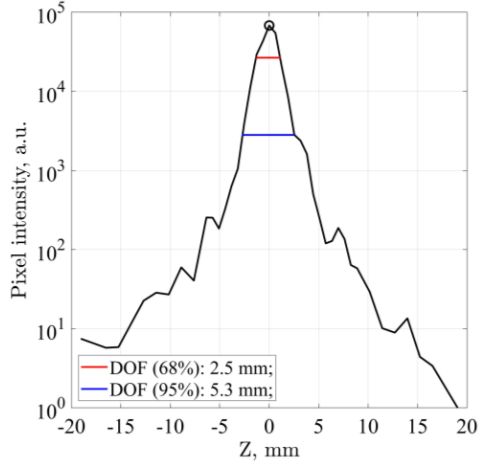

FIG. 8. Confocal optics response function, with defined DOF lines. Blue line corresponds to a region with 95% intensity and red line corresponds to a region with 68% intensity.

## IV. RESULTS

Measurements were performed in eight locations across the distance of several cm with focal distance of 300mm. Focusing optics is installed on the translational stage with micrometer, which allows for precise location control. Typical IVDF profile obtained with confocal and conventional LIF setups at $Z = 0.5$ is shown in FIG. 9. Both signals are normalized to maximum intensity. Signal-to-noise ratio was defined as SNR = $\mu_{signal}/\sigma_{sigmal}$, where $\mu_{signal}$ is a average signal value and $\sigma_{sigmal}$ is the noise standard deviation found after subtraction of the fitted function from the signal. For both methods the obtained SNR values are very similar: SNR = 7 and SNR = 90 for conventional and confocal setups respectively. Ion temperatures and mean velocities were determined by fitting the IVDF profiles with

$$f(v) = \left(\frac{M_i}{2\pi k_B T_i}\right)^{1/2} e^{-\frac{M_i(v-v_0)^2}{2k_B T_i}}, \tag{2}$$

where $M_i$ is ion mass, $k_B$ is Boltzmann constant, $T_i$ is ion temperature, and $v_0$ is the mean velocity. Error bars for ion temperature is defined as a standard deviation between three measurements. For ion velocity error bar is a combination of standard deviation of three measurements and wavemeter uncertainty, which is ~60 m/s. One can see that both methods result in very similar shapes of VDFs and close values for mean velocities and ion temperatures. The shift between VDFs (~ 0.1 km/s for the maximum LIF signal) can be attributed to the nonlocality of confocal measurements.

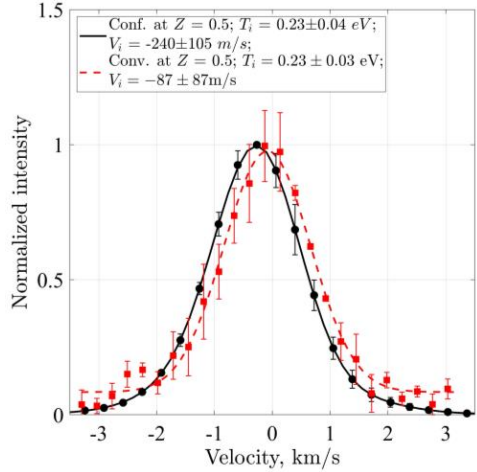

FIG. 9. Argon Ion Velocity distribution functions (VDF) at Z = 0.5 obtained with conventional (red, squares) and confocal (black, circles) LIF setups.

Mean ion velocities and ion temperatures along Z direction were compared with conventional LIF measurements performed at 3 positions, which are shown in FIG. 1. Results are shown in FIG. 10. One can see that overall flow velocities between two methods agree within the uncertainty. Ion temperature measurements shows good agreement as well. General trends for both quantities are similar between confocal and conventional LIF measurements.

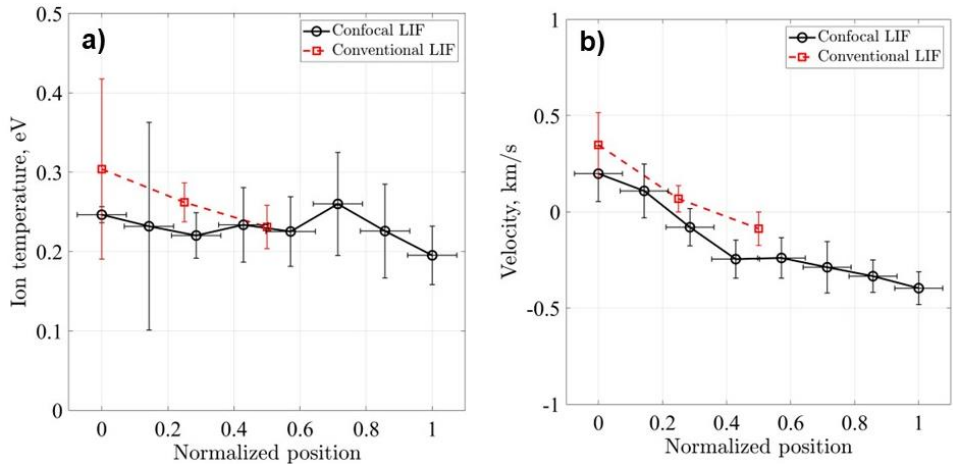

FIG. 10. Comparison of ion velocity (a) and ion temperature (b) distributions along the radial position obtained with conventional (dashed line) and confocal (solid) LIF setups. Vertical error bars are standard deviations from three repeated measurements. Horizontal errorbars correspond to spatial resolution defined in FIG. 7.

Integration of ion VDF obtained at different spatial locations will yield the metastable ion density profile. This profile was compared with ion density profiled measured with the biased Langmuir probe along the beam path. Normalized profiles are shown in FIG. 11. As one can see, LIF profile (black circles) is wider than density profile obtained with the probe measurements (req squares). This is due nonlocality of the confocal measurements and differences between metastable ion density and total ion density, measured by the probe. The effect of nonlocality can be eliminated if the optics response function is known (shown in FIG. 8). The reconstructed profile, obtained as a deconvolution of the LIF profile with the normalized optics response function, is shown as a black dashed line in FIG. 11. Deconvolution is performed with a custom MATLAB function [36]. It can be seen that the reconstructed profile closely follows the one measured with the probe.

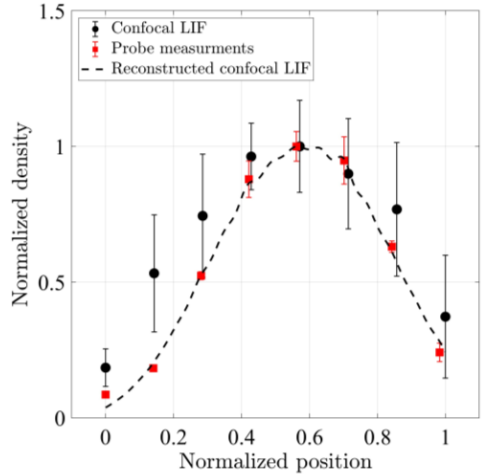

FIG. 11. Comparison of metastable density profile obtained with the confocal LIF (black circles) with the ion density measurements obtained with the biased Langmuir probe (red squares). Black dashed line shows density profile reconstructed from the deconvolution of the confocal LIF profile with the optical response function (FIG. 8). Error bars are standard deviations from three repeated measurements.

## V. CONCLUSION

In this paper, the confocal LIF configuration based on a ring-shaped laser beam was introduced and characterized both with and without plasma. The key element of this setup is the diffractive axicon optics, which ensures high ring beam quality, which is essential for achieving high spatial localization. The spatial resolution of the current setup is 5.3 mm. This is comparable to other reported confocal schemes at comparable focal distances [4]. However, the proposed design has some advantages. First, the spatial resolution is controlled by the laser beam parameters: ring thickness and beam diameter. It was shown that resolution of about 1 mm can be achieved with the same focal distance. At the same time, all fluorescence light, enclosed by the laser beam cone is collected. This allowed to maximize the SNR of the proposed LIF configuration. Finally, this LIF configuration avoids a problem with a beam back reflection, as the reflected beam from the back wall (if present) will diverge out of the field of view.

In plasma experiments, the proposed LIF configuration was utilized for measurements of argon ion VDF in an enclosed DC plasma source with a limited optical access. Comparison of the confocal and conventional LIF showed good agreement between determined plasma parameters (ion temperature and flow velocities). Ion temperature, found with conventional LIF, varied from $0.30 \pm 0.11$ eV to $0.2 \pm 0.03$ eV between $Z = 0 - 0.5$. Confocal LIF showed changes of ion temperature from $0.25 \pm 0.02$ eV to $0.22 \pm 0.05$ eV for the same range. Measurements of flow velocities with conventional method showed velocity change from $350 \pm 168$ m/s to $-90 \pm 87$ m/s in $Z = 0 - 0.5$. Corresponding flow velocity changes determined with confocal LIF were from $200 \pm 155$ m/s to $-240 \pm 100$ m/s for the same range. Reconstructed metastable density profiles extracted from LIF spectra were compared with the Langmuir probe ion density measurements. It was shown that density profiles show satisfactory agreement, which serves as a verification of the determined spatial resolution.

## VI. AUTHOR'S CONTRIBUTION

All authors contributed equally to this work.

## VII. CONFLICT OF INTEREST

The authors have no conflicts to disclose.

## VIII. ACKNOWLEDGEMENTS

This work was performed under the U.S. Department of Energy through contract DE-AC02-09CH11466. Authors would like to acknowledge Dr. S. Kondeti and Dr. S. Yatom for discussions and their help with the setup.

## IX. DATA AVAILABILITY

The data that support the findings of this study are available from the corresponding author upon reasonable request.